\documentclass[a4paper,11pt]{article}
\usepackage{pos}

\title{Fine grinding localized updates via gauge equivariant flows in the 2D Schwinger model}

\author*[a]{Jacob Finkenrath}

\affiliation[a]{Bergische Universit\"at Wuppertal, Gaussstrasse 20, 42119 Wuppertal, Germany }
\emailAdd{finkenrath@uni-wuppertal.de}

\abstract{ .}

\FullConference{%
 The 40th International Symposium on Lattice Field Theory, LATTICE2023
  31st July - 4th August, 2023
Fermilab
}

\abstract{State-of-the-art simulations of discrete gauge theories are based on Markov chains with local changes in the field space, which however at very fine lattice spacings are notoriously difficult due to separated topological sectors of the gauge field. Hybrid Monte Carlo (HMC) algorithms, which are very efficient at coarser lattice spacings, suffer from increasing autocorrelation times. This makes simulation of lattice QCD close to the continuum infeasible even with exa-scale computing.

An approach, which can overcome long autocorrelation times, is based on trivializing maps, where a new gauge proposal is given by mapping a configuration from a trivial space to the target one, distributed via the associated Boltzmann factor. Using gauge equivariant coupling layers, the map can be approximated via machine learning techniques. However the deviations scale with the volume in case of local theories and extensive distributions, rendering a global update unfeasible for realistic box sizes.

In this proceeding, we will discuss the potential of localized updates in case of the 2D Schwinger Model. Using gauge-equivariant flow maps, a local update can be fine grained towards finer lattice spacing. Based on this we will present results on simulating the 2D Schwinger Model with dynamical Nf=2 Wilson fermions at fine lattice spacings using scalable global correction steps and compare the performance to the HMC. }

\begin{document}
\maketitle

\section{Motivation}

Monte Carlo simulations in gauge theories show critical slowing down towards fine lattice spacings, i.e.~autocorrelations increases proportional to a higher inverse power of the lattice spacings \cite{Schaefer:2010hu}.
Especially in gauge theories, this is even more severe due to topological freezing.
Due to that it is currently not feasible to simulate with state-of-the-art algorithms at very fine lattice spacings below $a<0.04 \;\textrm{fm}$ with periodic boundary conditions, see for a review \cite{Finkenrath:2023sjg}.
However, if possible, simulations at very fine lattice spacings have a major impact on measurements in lattice QCD, namely by stabilising continuum extrapolations which would enable a robust way to make advances at the precision frontier of the standard model of particle physics.

In this proceedings we will investigate possible algorithmic solutions in case of the two-dimensional (2D) Schwinger model with two mass degenerated quarks. For the pure gauge action $S_g$ the Wilson plaquette definition and for the fermion action $S_f$ the plain Wilson fermion discretization $D(U)$ is used. The  Boltzmann weight or here also called target distribution is given by
\begin{equation}
 \rho(U) = Z^{-1} e^{-S_f(U) -\beta S_g(U) } = Z^{-1} \textrm{det} D^2(U) \cdot \textrm{exp}\{ - \beta \sum_{x=1}^V P_{12}(U,x)\}
\end{equation}
with $Z$ the partition function and $P_{12}(U,x)$ the plaquette at lattice point $x$.
The Schwinger model shares several properties with the 4D-QCD model, e.g.~the Hybrid Monte Carlo algorithm suffers from topological freezing towards larger values of $\beta$.
It also simplifies implementations and computations, due to that it is often used as a test bed to develop algorithms, as we will do here.

Simulations in lattice QCD are based on Markov Chain Monte Carlo (MCMC) algorithms, which generate an ensemble of configurations distributed via the Boltzmann weight $\rho(U)$. In general the MCMC procedure consists of two basic steps.
In the first step a new configuration $U'$ is proposed via a proposal probability
 $T_0 (U \rightarrow U')$
where we denote $\tilde{\rho}(U')$ as the probability distribution of the proposal.
In the second step the weight is corrected via an accept-reject step
\begin{equation}
P_{acc} (U\rightarrow U') = \textrm{min} \left[ 1, \frac{ \tilde{\rho}(U) \rho(U')  }{  \rho(U)  \tilde{\rho}(U')}\right]~.
\end{equation}
Now, the MCMC method works efficiently, if the proposal can decouple the proposed configuration $U'$ effectively from starting configuration $U$. The decoupling time is measured as the autocorrelation time, where the longest is usually associated with the topological charge in gauge theories. Thus, decoupling requires to have a high topological tunnelling rate for the proposal at a relative high acceptance rate in the accept-reject step. 

For the acceptance rate follows \cite{Creutz:1988wv}
\begin{equation}
P_{acc} = \textrm{erfc} \{ \sqrt{\sigma^2(\Delta S)/8 }\}
\label{eq:Pacc}
\end{equation}
with $\Delta S = \textrm{ln}(\rho(U')) - \textrm{ln}(\rho(U)) +\textrm{ln}(\tilde{\rho}(U))-\textrm{ln}\tilde{\rho}(U'))$
if the distribution $(\tilde{\rho}(U) \rho(U')) / (\rho(U)  \tilde{\rho}(U'))$ is log-normal distributed.

To reach high acceptance rate, the variance of $\Delta S$ has to be under control. This can be done by
\begin{itemize}
\item[1.] Using correlations between the proposal and target distribution, e.g.~by maximizing the covariance between $\rho$ and $\tilde{\rho}$.
\item[2.] Reducing the degrees of freedom within the proposal distribution, e.g.~by localization of updates via domain decomposition techniques.
\end{itemize}
Both methods can be combined using hierarchical filter steps as outlined in \cite{Finkenrath:2012az} to reach an even higher acceptance rate.

Now, in order to have a high tunneling rate of the topological charge, the proposal for a new gauge configuration should allow for topological transitions. One possible choice of such proposal is given by trivializing flows \cite{Luscher:2009eq}. The basic idea is to start from a configuration weighted via a uniform distribution $r(U_0)$ and to define a map $U=f^{-1}(U_0)$, which transforms the configuration to the target space distributed by $\rho(U)$. Using an appropriate definition of the map the distribution of the flow $\tilde{\rho}$ can be calculated via the Jacobian of the transformation, which is given by
\begin{equation}
 \tilde{\rho} (U) = r(f(U)) \cdot \left| \textrm{det} \frac{\partial f(U)}{\partial U} \right|.
\end{equation}

Examples for a tractable maps are given by normalizing discrete flows  \cite{Albergo:2019eim,Kanwar:2020xzo,Boyda:2020hsi,Albergo:2021vyo,Abbott:2023thq} or continuous flows \cite{Bacchio:2022vje}, see also \cite{Kanwar:2013lat}.
Here, we will use the former one \cite{Albergo:2021vyo}, where the flow distribution can be trained using convolutional neural networks within the map transformations also called coupling layers.
The training or optimization of the networks is efficient if the acceptance rate of the accept-reject step is high or if the corresponding variance $\sigma^2(\Delta S)$ is under control, compare eq.~\eqref{eq:Pacc} . This is usually achieved by using the KL-divergence which leads to $\tilde{\rho}(U) \approx \rho(U)$ \cite{Kanwar:2020xzo}.

Note the variance $\sigma^2(\Delta S)$ is an extensive quantity, i.e.~the lattice action is a function of the volume.
In case of a local theory, where the correlation length decays with $\textrm{cov}(x,y)\propto e^{-d |x-y|}$ and $d>0$, the variance of an extensive quantity, like the plaquette action, can be written as  $\sigma^2 = V (a_0 + a_1 e^{-d} + a_2 e^{-d \sqrt{2}} + \ldots )$ with $a_n$ the coefficient of the $\sqrt{n}$ distance suppressed via $e^{-d \sqrt{n}}$ and $V$ the volume. Now, 
by training the coupling layers the target distribution is approximated to a certain degree.
If we fix this approximation to $S$, the difference to the target distribution will scale with the volume. This means that to achieve a reduced scaling behavior the approximation needs to be improved when increasing the volume. This will become very hard or computational very challenging if higher order terms or very small long range correlations are left and are required to be learned or approximated to further improve the model. 
Due to that larger volumes with high acceptance rate are currently out of reach \cite{Kanwar:2013lat,DelDebbio:2021qwf,Abbott:2022zsh} and will require further advances in the definitions of the maps as well as in the training procedures.

\section{Localization}

\begin{figure}
 \centering
\includegraphics[width=0.54\textwidth]{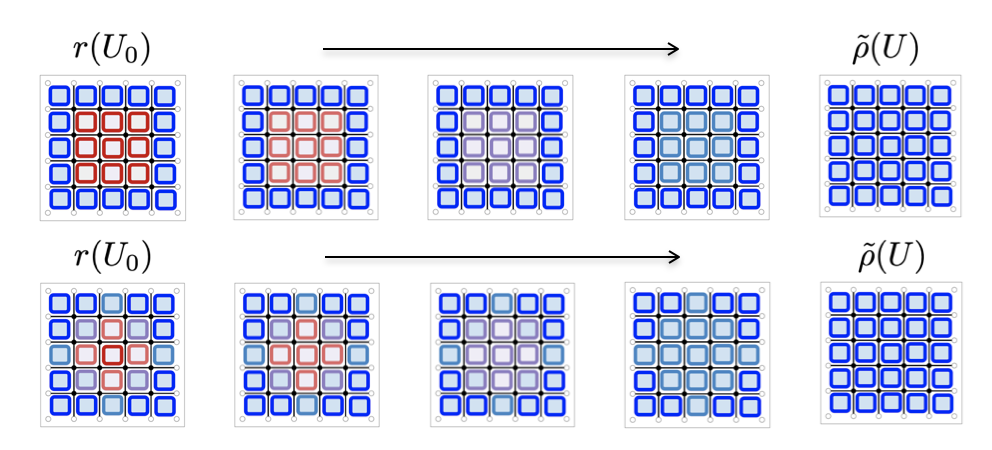}
\includegraphics[width=0.43\textwidth]{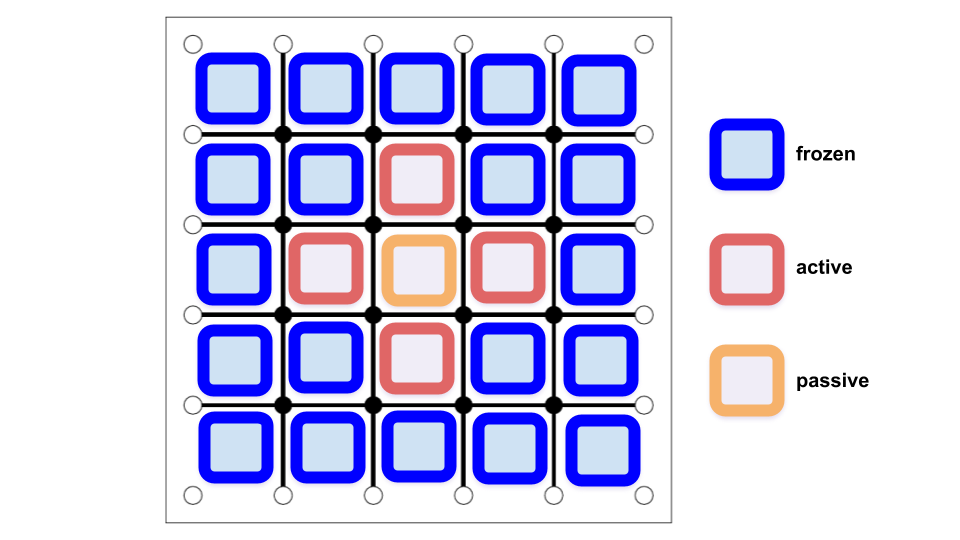}
\caption{The left panel illustrates the flow maps which transforms the plaquettes from a uniform distribution (red) towards the target distribtution (blue). The upper case shows the approach introduced \cite{Finkenrath:2022ogg}, where the plaquettes and with it the corresponding links in the boundaries are fixed. The lower case shows the here introduced fine graining maps, where only the central plaquette is updated and smoothed out including the neighbouring plaquettes towards the target distribution. The right panel shows the basic kernel used to update the active plaquettes (red) involving frozen ones (blue) but not passive ones (organge) within an application of a coupling layer.  }
\label{fig:maps}
\end{figure}

A natural way out of volume scaling in local theories is the localization of the update. This can be done by decomposing the lattice into domains. 
Now, we can freeze the boundary links of each domain, which share loops with neighbouring domains, and only update links within each domain. 
This can be applied to the equivariant flow map, as illustrated in the left panel of Fig.~\ref{fig:maps}.
While some modifications compared to the periodic case in the training are needed, the approach works well for lattice sizes of $L=8$, as demonstrated in \cite{Finkenrath:2022ogg}.
However, for larger volumes updates become difficult due to the volume scaling, which makes a direct adaptation to 4D-SU(3) without improvements difficult.

\subsection{Fine graining flows}

Before we discuss the here introduced modification to the localized flow, we want to point out an observation how flow maps are constructed. 
The standard procedure is based on a map at fix number of lattice points or at fixed lattice extents.
In case of a trivializing flow, which is reducing effectively the lattice spacing during mapping, the physical size of the box is changing. This implies that during the mapping to the target theory, the projection has to filter out un-wanted noisy modes and need to build up longer correlations at the same time to match to the physics at the target space.

A more natural projection is given by keeping the physical box size fixed during the mapping. This however would imply for a trivializing map to add new physical degrees of freedoms to new intermediate lattice points in the discrete lattice during a mapping from a coarse to a finer theory. While with machine learning techniques this might be in range~\cite{Abbot:lattice2023},
such operation requires additional non-smooth steps which are non-trivial to tune.

Here, instead, we will use an effective fine graining approach, by introducing a map which smoothes out a local defect. This is done by placing a local defect  into a larger lattice or domain and fine grain it into the local neighbourhood. 
A similar approach is done using multi-tempering, which is successfully applied in unfreezing topological charges in 4D-SU(N) pure gauge models, see ~\cite{Bonanno:2020hht}.

Here, we will use local flow transformations to map between the target distribution and the defect, which is generated by updating the links within the defect according to a uniform distribution.  
This can be also applied within a multi-tempering algorithm, like proposed in the T-REX algorithm for 4D-SU(3) in \cite{Hackett:lattice2023}. In this proceeding, we will use a uniform update of the defect to enable topological transitions. This will require to adapt the training procedure to find an appropriate map.

\begin{figure}
 \centering
\includegraphics[width=0.75\textwidth]{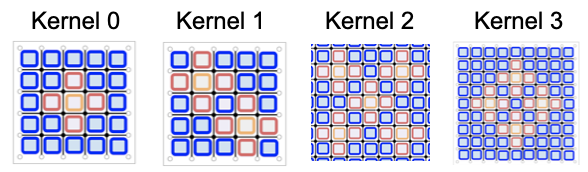}
\caption{The figure shows the kernels for the maps used for the different coupling layers.
Each link of a passive plaquettes (orange) is associated with an active plaquettes (red) to achieve a compact layout.
The frozen plaquette (blue) are fixed and used as input to update the active plaquettes.}
\label{fig:kernels}
\end{figure}

 We will follow~\cite{Kanwar:2020xzo,Boyda:2020hsi} by using the gauge equivariant flow. For a local defect, we introduce a localized map-condition to update the plaquettes. Namely, we use in 2D a maximal compact map by placing the passive plaquette in the center and updating all links belonging to it by using the neighbour plaquettes. This gives an active to passive plaquette ratio of 4:1, see right panel of Fig.~\ref{fig:maps}. Note, that in principle for a unit N-D cube in higher dimensions N the ratio decreases, however additional passive links outside the cube can be included using the procedure outlined in \cite{Boyda:lattice23}.
 Now, we can design additional center symmetric kernels by placing the introduced symmetric kernel around the defect, as illustrated in the left panel of Fig.~\ref{fig:kernels}.
Additional we add a smoothing kernel, which we denote as the 4th kernel, by using nine times the 2nd kernel and shifting with $\{(1,0),(0,1),(-1,0),(0,-1),(1,1),(1,-1),(-1,-1),(-1,1),(0,0)\}$ respectively.

 To ensure that the proposals allows topological tunneling, we modified the training conditions by only using sets from the batch for the optimization process of the coupling layers, for which the topology changed. Namely, by introducing a \textit{topological} loss function, given by
 \begin{equation}
 L = \textrm{ln} ( \tilde{\rho}(U') / \rho(U') ) \cdot | Q(U') - Q(U) |
 \end{equation}
 with $Q$ the topological charge, which is exactly quantified in the 2D-U(1) model. Note, if the standard loss-function is used
the flow would become most likely a \textit{fancy} plaquette update.

\begin{figure}
 \centering
\includegraphics[width=0.75\textwidth]{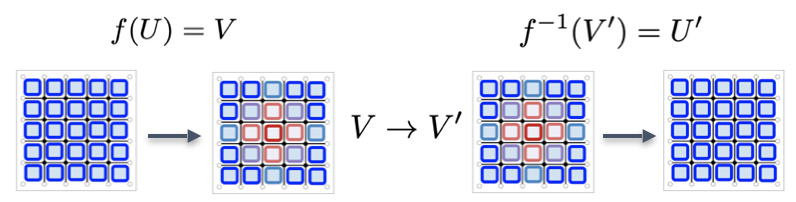}
\caption{The figure illustrate the fine graining update step. In the first step the reverse mapping is applied to the starting configuration.
Then, the links of the inner plaquette (red) are updated using a uniform distribution. In the final step the plaquettes are mapped back towards the target distribution.}
\label{fig:update}
\end{figure}

Fine graining the defect to the neighbouring links requires to adapt the proposal and accept-reject procedure of the MCMC chain. Namely, it requires to first apply a back transformation to the configuration, before the defect can be updated, see Fig.~\ref{fig:update}. To stabilize the training of the coupling layers the back-transformation is fixed and updated after several epochs. In general, this allows to introduce an additional loss-function given by
 \begin{equation}
  L = \textrm{ln}\{\rho(U') \} - \textrm{ln}\{\rho(U)\} + \textrm{ln}\{ \tilde{\rho}_{\textrm{fixed}}(U) \} - \textrm{ln}\{ \tilde{\rho}_{\textrm{train}}(U') \} 
 \end{equation}
 which is used additional to the topological loss-function to increase the acceptance rate.
 
The training for a flow map to increase topological tunneling requires a fine tuning or grinding process. 
Namely, first the maps are pre-trained on a $L=8$ lattice with periodical boundary conditions using the topological loss. Note, if we use the 4th kernel plaquettes within the pretraining also the boundary plaquettes are updated. 
This map is then used as the fix backwards transformation on a $L=16$ lattice with fixed boundary condition, i.e.~the flow acts only on the center links close to the placed defect. The training is then iterated by updating the fixed backward maps and the procedure is further fine tuned by using smaller optimization steps.

\section{Results}
\subsection{Fine graining updates}

  \begin{figure}
 \centering
\includegraphics[width=0.44\textwidth]{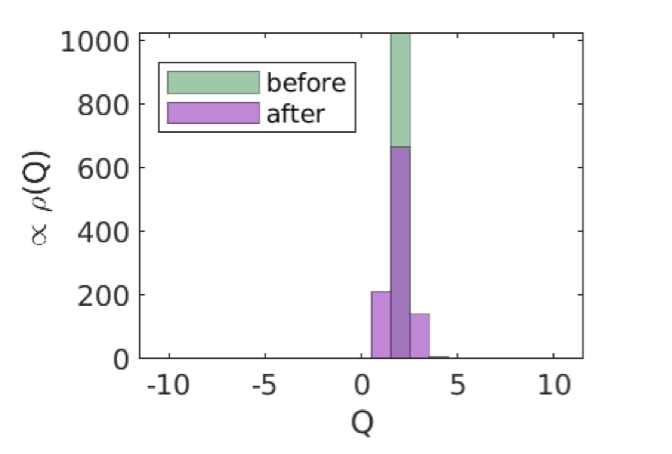}
\includegraphics[width=0.34\textwidth]{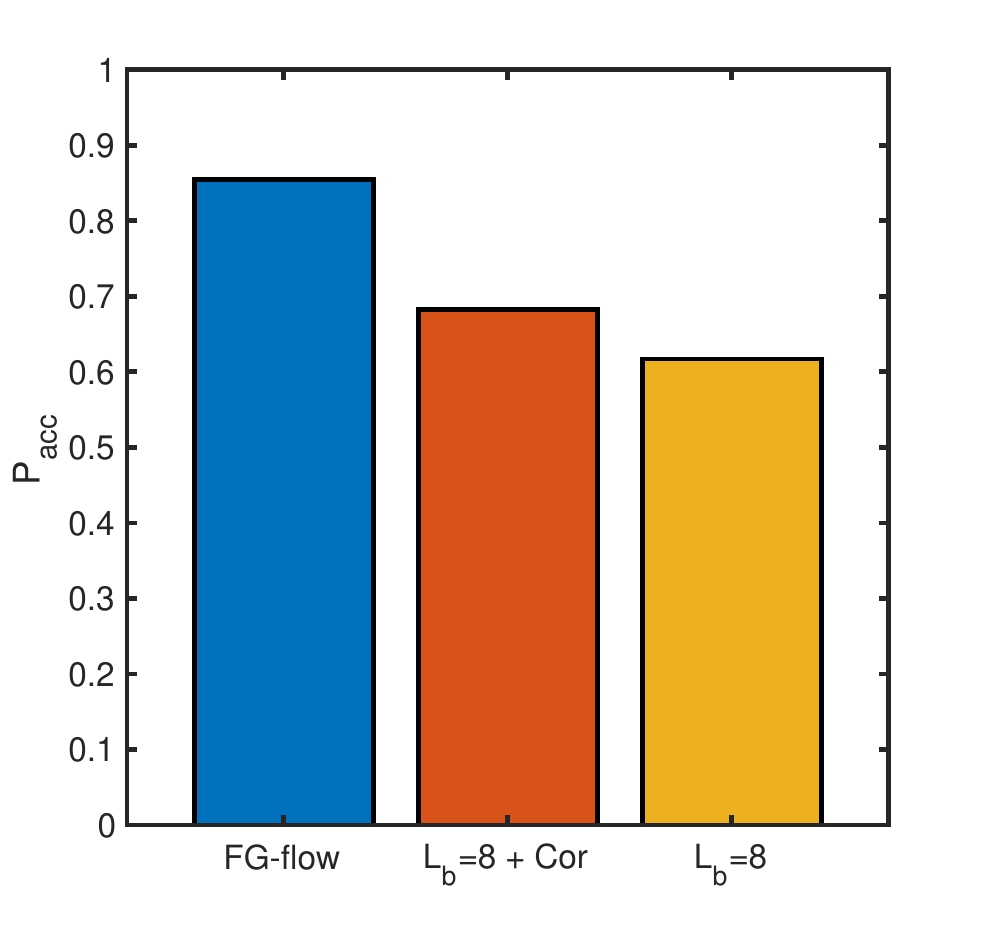}
\caption{The left panel illustrates the rate of a topological transition of a fine graining flow update on a fixed $L=16$ lattice at $\beta=11.25$.
For that, the fine graining flow update is repeated 1k times with the same initial configuration.
The right panel shows the acceptance rate of the local block determinants at $\beta=8.45$ and $m=-0.025$. The block size of the fine graining updates are given by $V=16\times16$ (blue). This is compared to the acceptance rate of the localized flow update  \cite{Finkenrath:2022ogg} 
using a block size of $V=8\times8$ including correlation with the pure gauge action (orange) and without (yellow).}
\label{fig:qrate}
\end{figure}

An example for a successfully trained flow map at $\beta=11.25$ is illustrated in the left panel of  Fig.~\ref{fig:qrate}
in case of the topological charge distribution on a single configuration before and after the fine graining update.
For the fine graining flow map, we used a combination of the kernels $[1,3,2,1,3,2,1,3,2,4]$, resulting into 18 coupling layers in total. 

Compare to the fixed boundary update, introduced in  \cite{Finkenrath:2022ogg}, the FG-flow updates shows also a higher acceptance rate in case of local corrections via the associated block determinant. 
This is shown in the right panel of Fig.~\ref{fig:qrate}, where the fine grain (FG)-flow update is done within a $16\times 16$ block and reached an acceptance rate of $85\%$ while the fixed boundary flow only reached $65 \%$ acceptance rate using a $8\times 8$ domain.

\subsection{Topological transitions }

\begin{figure}
 \centering
\includegraphics[width=0.52\textwidth]{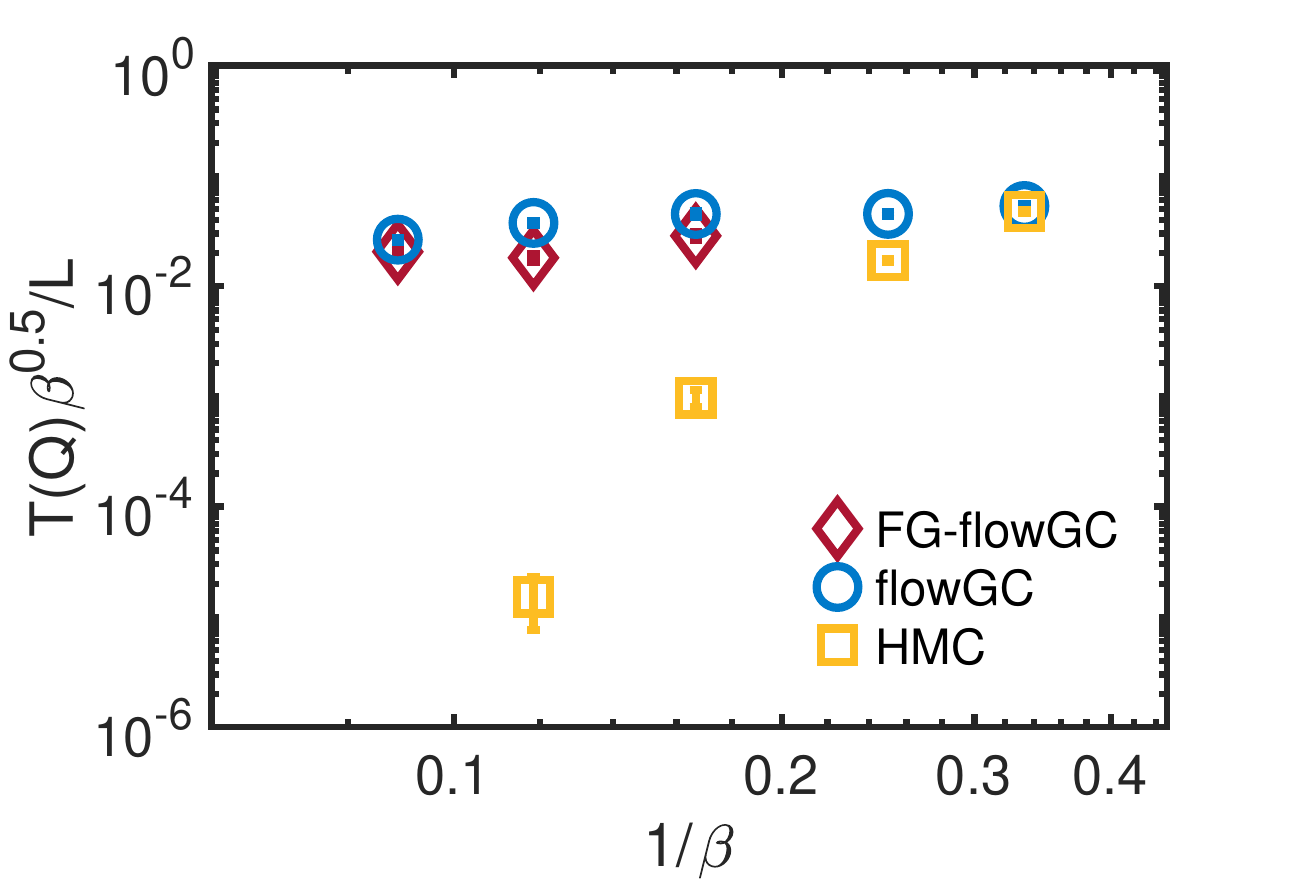}
\includegraphics[width=0.47\textwidth]{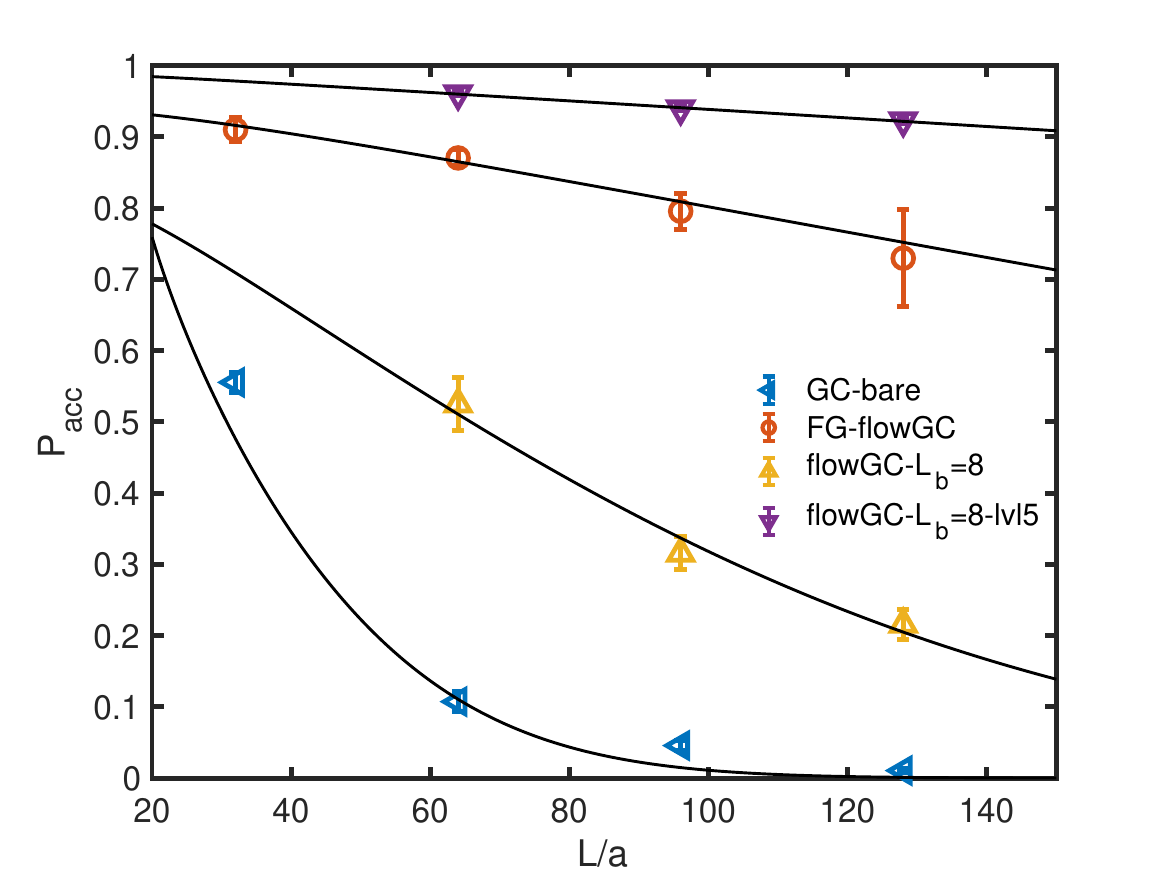}
\caption{The topological tunneling rate normalized with $\sqrt{\beta}/L$ is shown on the left panel obtained for the HMC (yellow squares), for the flowGC algorithm (blue circles) and the FG-flowGC at fixed lattice size $L/a=64$ (red diamonds). The right panel shows the volume dependence of global acceptance rate for the different global correction step algorithms at a gauge coupling of $\beta=6.0$ and a mass scale of $m_{PS} \sqrt{\beta}=0.4$, see \cite{Christian:2005yp}. The impact of using hierarchical filter are demonstrated, i.e.~the acceptance rate using the full determinant drops fast by increasing the lattice size (blue (left-pointing) triangles), using a 4-level method with domain decomposition with $L=8$ blocks increases the rate (yellow (top-pointing) triangles) and using a 5-level method with recursive domain decomposition increases the rate (purple (down-pointing) triangles)) even further, see also \cite{Finkenrath:2022ogg}. The , here introduced, flow grain update is using a 4-level methods with domain decomposition of $L=16$ blocks (orange circles). }
\label{fig:trate}
\end{figure}

Now, we can use the FG-flow updates within a MCMC procedure. To achieve an ergodic algorithm, the FG-flow updates are combined with a Hybrid Monte Carlo (HMC) update step, i.e.~after each FG-flow update a shift and a trajectory via the Hybrid Monte Carlo algorithm is performed. To include the fermion weight hierarchical filter steps in combination with the domain decomposed determinant like discussed in  \cite{Finkenrath:2012az,Finkenrath:2022ogg} are used, which we will denote as flowGC. The final step is given by a global correction step (GC) with the determinants of the corresponding Schur complements \cite{Finkenrath:2012az}.
The normalized topological transition rate $T(Q)\sqrt{\beta}/L$ with $T(Q) = \langle | Q_i - Q_{i+1}| \rangle$ is used in a range of $\beta \in [3.0; \, 11.25]$ as shown in Fig.~\ref{fig:trate}.
The comparison between flowGC and FG-flowGC is done at $L/a = 64$ while for the HMC the rate is measured by fixing $L/\beta \approx 40$ and a mass-scale set to $m_{PS} \sqrt{\beta}=0.4$, where $m_{PS}$ is the pseudoscalar mass, see \cite{Christian:2005yp}. By only randomizing $0.8\%$ of the links the FG-flow procedure shows a similar transition behaviour like the flowGC, which both unfreeze the topological charge successfully in contrast to the pure gauge HMC algorithm.

\subsection{Comparison with other global correction approaches}

\begin{figure}
\centering
\includegraphics[width=0.75\textwidth]{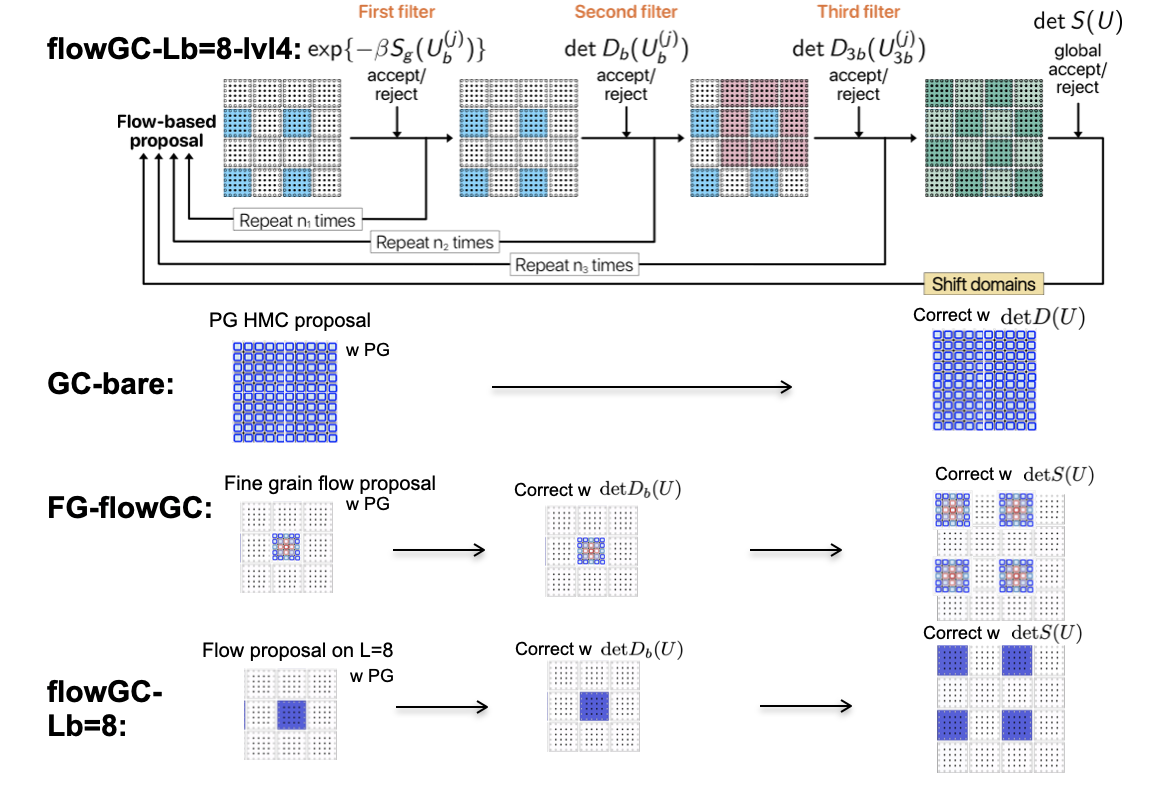}
\caption{The sketch illustrates the different filter steps of the global correction step algorithms. While the 5-level approach, discussed in \cite{Finkenrath:2022ogg}, using two filter steps including block determinants on $L=8$ and $L=24$, the 4-level flowGC only uses the $L=8$ blocks and the \textit{GC-bare} only includes the global determinant. The here used FG-flowGC algorithm is based on a 4-level approach, using block determinants of $L=16$ as a intermediate filter step.}
\label{fig:arate}
\end{figure}

The FG-flowGC has not only a high acceptance rate for the correction of the local block determinants, it shows also good performance at the global correction step. This is illustrated in Fig.~\ref{fig:arate}.
The different procedures are given as follows:

\begin{itemize}
\item \textbf{GC-bare} with two-levels 
\begin{itemize}
\item pure gauge update via HMC of the full lattice
\item global correct reject step including correlations 
\end{itemize}

\item flowGC with $L=8$ and 4 levels:
\begin{itemize}
\item  flow update within domains by fixing each boundary links (updating only every 4th domain)
\item accept reject with local pure action (second level), $L=8$ block determinants (third level)
\item  global correction step with determinant of the Schur complement
\end{itemize}

\item  flowGC with $L=8$ and 5 levels:
\begin{itemize}
\item  similar to flowGC, only adding an additional filter level before the global correction step, by introducing an extended domain with $L=24$ including each of the updated domains
\end{itemize}

\item FG-flowGC with 4 levels:
fine graining updates within a $L=16$ domain
\begin{itemize}
\item  similar to the flowGC with 4 level only using fine graining updates within $L=16$ domains.
\end{itemize}
\end{itemize}

Note that all filter steps before the global correction steps are local and can be perform in parallel, making the approach scalable and inline with current algorithmic trends such as highlighted in \cite{Finkenrath:2023sjg,Boyle:2013lat}.
Moreover, the approach can be easily generalized to the multi-level procedure  \cite{Ce:2016ajy,Ce:2016idq}, which would not only help to fight against long autocorrelation times but also mitigate exponential increasing noise in the long-distance measurements of correlators. Additional the procedure can be combined with other updates schemes, which show success in pure gauge simulations such as multi-tempering approaches \cite{Bonanno:2020hht,Eichhorn:2023uge}.

\section{Conclusion}

Within this proceeding we discussed a new way to design flow maps to decouple the topological charge by updating only a minimal number of links in the 2D Schwinger model.
The trained flow are given by fine graining updates, which are updating a local defect using center symmetric kernels.
The current status of the training procedure involves several steps resulting in a fine tuning procedure or a grinding task. However, using the fine graining updates high acceptance rates are achieved for the fermion corrections, for the local block determinants as well as for the global correction steps. Furthermore the update unfreezes topology, i.e.~thus solve the key problem. If the training procedure is under better control, the next step is its application to 4D-SU(3).
\vspace{0.4cm}

\textbf{Acknowledgments:}  
J.F.~received financial support by the German Research Foundation (DFG) research unit FOR5269 "Future methods for studying confined gluons in QCD".  The author gratefully acknowledges the Cyprus Institute for providing computational resources on Cyclone for training the neural networks.

\end{document}